\documentstyle[11pt]{article}
\def\alt{\;{\lower 1.5pt\hbox{$<$} \above 0pt \raise 1.5pt\hbox{$\sim$}}\;}
\def\agt{\;{\lower 1.5pt\hbox{$>$} \above 0pt \raise 1.5pt\hbox{$\sim$}}\;}
\def\notl{\ell\kern-4.5pt\hbox{$/$}\ }
\def\bard{d\kern-2.5pt\raise 3pt\hbox{-}}

\begin{document}
\vbox{
\begin{flushright}
AZPH-TH/93-07
\end{flushright}
\title{A Renormalization Group Approach to the Chiral Transition}
\author{Arjun Berera* \\
Department of Physics \\
University of Arizona \\
Tucson, Arizona 85721}
\date{}
\maketitle
\begin{abstract}
A methodology is given to test the QCD $N_f$=2 chiral transition, presently
conjectured to be second order.  Scaling forms for the correlation
length, susceptibilities and equation of state are given which
account for finite lattice spacing.  Confirmation by lattice
simulation would provide a large set of consistency checks for
establishing that the transition is second order.  Further
corrections from finite volume effects and higher dimensional
operator mixing are given.  The implications
of scaling corrections
in finite temperature lattice QCD studies are examined with
emphasis on
tests for the believed second
order chiral transition within the realistic setting
of finite lattice spacing effects.
\end{abstract}
\vspace{20mm}

\bigskip
}
\eject

Arjun Berera \\
Department of Physics \\
University of Arizona \\
Tucson, Arizona 85721 \\
phone:(602) 621-6790 \\
FAX: (602) 621-4721 \\
e-mail: aberera@corelli.physics.arizona.edu \\

Running head: RG Approach to Chiral Transition \\

\eject
\noindent {\Large\bf Section I. Introduction}
\medskip

Landau-Ginzburg-Wilson(LGW)-Hamiltonians have played a central
role in the order parameter, renormalization group
approach to second order phase transitions.
Their importance to QCD was realized by Pisarski and
Wilczek \cite{pisarski}, who
pointed out that the QCD chiral phase transition has
a suitable order parameter from which an effective
critical theory could be constructed. By
following standard notions of universality, they accounted for
all the symmetries of QCD in the chiral limit in order to build an
order parameter based effective theory.  This led them to the class of
SU(n)$\times$SU(n) $\sigma$-models.  Then, according to the
universality hypothesis, it
follows that
these effective theories would give the same scaling behavior
as the true microscopic theory, here QCD,
near the critical point, provided that their order parameter ansatz
is correct.

Recently Wilczek \cite{wilczek} has examined the case of two fermion flavors
in the chiral limit
where the effective LGW-theory is now the
SU(2)$\times$SU(2)=O(4), Heisenberg model.  Since
the latter theory has been well studied in statistical
mechanics, the necessary information regarding
the critical behavior of this theory was immediately available.
Subsequently Rajagopal and Wilczek \cite {rajagopal}
have given a transcription dictionary
which relates the O(4)-Heisenberg transition to
the language of lattice gauge theory.
This transcription readily allows testing the original
assertion, that the chiral phase transition is
second order, by checking the universal scaling behavior
and comparing to theory.

This paper provides a theoretical consideration of
critical phenomena scaling in the context of
lattice gauge QCD.  Although the general theory of scaling
in critical phenomena is well studied, several
attributes specific to lattice gauge theory have not been
adequately appreciated.  It is our goal to provide some of the
missing ingredients here.  As one outcome of these considerations,
we hope it will motivate more extensive numerical
lattice studies near the transition or crossover region, so that
reliable scaling studies can be made.
For this, a reasonable sample of points
would be needed within a region no greater than
10 percent of the believed transition temperature.
For example for the
magnetization, $< \bar \psi \psi>$, where
simulation data is fairly reliable, presently there are
about three simulation points
within the fall-off portion of the magnetization profile.
With this data, at best one can approximate
the transition point.  If one wants to conduct a serious
renormalization-group scaling study, one needs
at least a factor of five increase
in data points in order to make reliable curve fits.
Irrespective of how motivated the reader is by this paper,
one can also turn to
condensed matter physics, where present wisdom considers
scaling studies as among the optimum tests for second order behavior.
A resolution of the QCD chiral transition could now
also benefit by intensifying quantitative efforts towards
this means.  Some efforts along these lines recently have been made
by Karsch \cite{karsch}.  He has examined scaling behavior of what he refers
to as the pseudo-critical coupling.

Theoretically, the considerations of \cite{wilczek,rajagopal} regarding
scaling are not complete since they overlook
lattice specific corrections to leading scaling.
Intrinsically, the effective theory in \cite{wilczek} represents only
a truncated sector of the complete QCD Hilbert space.  In
particular it does not include the full effect
of short distance quark and gluon degrees. Their presence
renormalizes the parameters of the effective
theory and induces irrelevant, higher dimensional, operator correction
terms to it.
This leads to intrinsic scaling
corrections to the critical behavior of QCD.
Furthermore, in a lattice simulation there are also
corrections to leading scaling which are extrinsic to continuum
QCD.  These effects arise from the lattice spacing, which at leading order
induces dimension four corrections.
Thus if one accepts the arguments of \cite{pisarski,wilczek,rajagopal},
or similar
such based on symmetry considerations, that the chiral transition
is second order, its critical behavior from a lattice simulation
will have corrections to leading scaling due to the presence
of the lattice spacing.
In the analysis of \cite{karsch} these effects were not
treated.

In total a lattice simulation involves four
scales which in ascending order are
the lattice spacing, the intrinsic cutoff $\Lambda_{QCD}$,
the correlation length
and the lattice volume.  Ideally, the intrinsic length should be
much bigger than the lattice spacing.  In this case
corrections associated with critical
phenomena would be independent of the lattice spacing and
only the continuum properties of QCD would be measured.
In practice, typical hadronic lengths are only
a few lattice spacings in size and are strongly dependent on
the lattice size.
Thus, it is important
to understand the effects of finite lattice spacing since they
reflect upon lattice artifacts that one is forced to live with
in any real simulation.

Following in the spirit of universality,
we will discuss in this paper those aspects of nonideal
lattice complications that induce universal effects. Since such
effects will invariably appear in real simulations, it is
useful to understand what we can learn from the well
developed renormalization group theory of scaling
and scaling corrections.
What we will see is that this approach provides a way to study the critical
properties near the chiral transition with rigorous
account for the lattice spacing.  As such it opens
the possibility for lattice studies to circumvent their inevitable
problems with lattice spacing effects and obtain further
answers regarding the chiral transition.
In addition to
offering a strong set of consistency checks regarding the
transition,
since lattice effects can be explicitly treated, the
methodology is also appropriate for further exploration
of this transition from data obtained by simulations.

The paper is organized as follows.  In section two,
as a brief refresher, we will review
the renormalization group theory of second order phase
transitions.
In section three, scaling forms are presented which
include lattice spacing correction effects.
Other corrections arising from finite volume effects and higher
dimension
operators are also
given for completeness.
Finally, we give our conclusions and discuss various
applications where scaling corrections will be of use
in analysis of simulation data.

\bigskip
\noindent{\Large\bf Section 2}
\medskip

In this section we will review the application of the
renormalization group to studies
of infra-red fixed points.
Imagine one is given a Hamiltonian which, due to universality, can be
any of a number of choices so long as it satisfies all the symmetry
properties and has the correct spatial
dimensionality as that of the original system of interest.  In this
respect, for our problem of the chiral phase transition
a suitable such Hamiltonian as stated above is the O(4)
Heisenberg model,
$$
H=\int
d^dx^\prime\left\{{1\over2}\sum_{a=1}^4\left[c\nabla\phi^\prime_a\cdot\nabla\phi^\prime_a
+m^\prime_0{\phi^\prime_a}^2\right]
+{\lambda^\prime\over4!}\left(\sum_{a=1}^4{\phi^\prime_a}^2\right)^2\right\}
\eqno(2-1)
$$
Recall that since this describes a statistical mechanical system,
implicitly there is a lattice cutoff $a=1/ \Lambda $.
We wrote the theory in continuum notation to keep the closest
analogy to field theory.  However when doing
momentum space calculations, recall that all integrals have an
ultraviolet cutoff.  Observe also that
in statistical mechanics the bare Hamiltonian
is the actual physical theory in contrast to quantum field theory
where the renormalized theory is the physical one.
The goal now is to study the universal critical behavior of
this theory using the tools of field theory.

The philosophy of the field theoretic approach to critical phenomena
is the same as the Wilson-Kadanoff blocking approach in that
one wants to coarse grain the original theory and generate
a set of effective Hamiltonians, each describing larger cells
of the original theory by way of thinning the dynamical degrees of
freedom.  To implement this idea field theoretically, one first
rescales spatial and spin coordinates to appropriate magnitudes.
In order to fix ideas, we will view the rescaling as
a redefinition of the coordinate system with the physical system
defined by the initial prime coordinates.  Thus a momenta $p$
for example in the system scaled by $ \Lambda $ as in (2-2) below
becomes $p/ \Lambda $ in the physical system.
{}From this point of view,
as one examines a correlation function, for example, at a fixed
momentum $p$ but in a succession of scaled coordinate systems, meaning
$ \Lambda \to \infty $, one is studying the $p \to0$ behavior
of the correlation function for the original system.
To define the scaled coordinate system $x$ in terms of the original
(physical) prime coordinate system $x^\prime$,
for
the spatial coordinates, since
typical distances near the critical point
are large, one defines
$$
x^\prime=\Lambda x \eqno(2-2)
$$
where $ \Lambda \gg 1$.  By this transformation, the relevant
magnitude of $x$ will now be of order one.  Based on engineering dimensions,
near the critical point the field variables are expected
to be small in magnitude.  As such one defines,
$$
\phi^\prime_a(x^\prime) = {\sqrt c\phi_a(x^\prime)\over\Lambda^{1-d/2}}
\eqno(2-3)
$$

One now substitutes these quantities into (2-1) to obtain
$$
H=\int d^dx\left\{{1\over2}\sum_{a=1}^4\left[\nabla\phi_a\cdot\nabla\phi_a
+m_0\phi_a^2\right]
+{u_0\over4!}\left(\sum_{a=1}^4\phi_a^2\right)^2\right\}
\eqno(2-4)
$$
where the bare mass $ m_0 $ is related to the temperature by
$ m_0 \sim T- T_0 $ with $ T_0 $ being the mean field
transition temperature.
The first stage of the coarse graining is realized here in the sense
that distances of order one in the unprimed coordinates
are large in the original primed system.  At an intuitive level
one can conceptualize the effect of this coarse graining as saying
that the
energy density sampled in a cell $dx$  in the scaled
system is $ \Lambda $ times
greater than in the original system.  In the context of the
Hamiltonian, the result of
this first stage of coarse graining can alternatively be viewed as that of
redefining the bare parameters of the unprimed system with respect
to the original primed system.  In particular the
four body interaction coupling
is,
$$
u_0={\lambda^\prime\Lambda^{4-d}\over c^2}=g_0\Lambda^{4-d}
\eqno(2-5)
$$
so becomes infinite as $ \Lambda\to \infty $ for $d<4$ (recall
that the relevant physical dimension of interest for
finite temperature QCD
is d=3).
Note here that the transformation acts only on the spatial and field
coordinates and not on the parameters which in particular
include the lattice spacing.  Thus if one examines a correlation function
at momentum p in the unprimed coordinate system scaled by
$\Lambda$ with lattice cutoff a, it is equivalent to the correlation
function in the original primed coordinate system at momentum
$p/ {\Lambda}$ but with the same lattice spacing a.
In practical terms this means one can compute the renormalization
properties of correlation functions in the limit
$\Lambda\to \infty$, as one is used to doing in field theory,
and at the same time be studying the infrared properties
of the correlation functions of interest.

To study critical phenomena with this formalism, one must
demand that the theory be independent of $ \Lambda $ in the
limit $ \Lambda \to \infty $.
Recall that in statistical
mechanics there is no physical reason for demanding the
above condition nor does it a priori lead to a scale invariant
solution.  It should be regarded as an ansatz for
obtaining potentially scale invariant solutions.
In this
respect, this field theoretic approach cannot predict the critical
temperature nor even determine a priori that a second order phase transition
exists.  The most it can do is assume the existence of one
and check a posteriori.  The latter is achieved if an infrared fixed point
is found which will be discussed further below.  Given that a fixed
point is found, then the power of the field theoretic approach
is its greater accessibility to analytic treatment compared to the
Wilson-Kadanoff
blocking approach.  The quantities that
can be computed are the scaling behaviors of
all correlation functions, which in particular
are seen explicitly to satisfy known scaling laws.  From this
critical exponents can be obtained, as well as the
equation of state.

How one defines the renormalization of the theory is not
specified nor unique.  In this respect, if a particular renormalization
procedure does not yield an infrared fixed point in perturbation theory,
it does not
mean the initial theory has no second order phase transition.
All one can conclude is that the sought for fixed point may require
a different renormalization procedure in order for it to be reached by
perturbation theory or that it may only be reached by nonperturbative
means.  On the other hand,
if a fixed point is found, one has reasonable confidence
that a second order phase transition exists in the system.

Our discussion of renormalization will follow the conventional
approach.  Here one aims to take
the $ \Lambda \to \infty $ limit while holding fixed a
certain minimal set of computed quantities.
The statement of renormalizability in this standard approach
is that one can adjust a finite number of parameters
to make all observables finite.
To renormalize the theory, we define
the renormalized correlation functions as,
$$
\Gamma^{(N)}(p,g_0,\Lambda)=Z^{-N/2}(g_{0}, {\Lambda\over\mu})
\Gamma_R^{(N)} (p,g,\mu,\Lambda)
\eqno(2-6)
$$
where $Z$ are the renormalization scaling parameters.
The next step
is to give the renormalization conditions which we
take as,
$$
\Gamma_R^{(2)}(p,g,\mu)|_{p^2=0}=0
$$
$$
{\partial \over \partial {p^2}}\Gamma_R^{(2)}(p,g,\mu|_{p^2=\mu^2}=1
\eqno(2-7)
$$
$$
\Gamma_R^{(4)}(p_i,g,\mu)|_{sp}=\mu^\epsilon g
$$
The $ \Lambda \to \infty $ limit can now be taken, and the renormalized
correlation functions will by construction be independent of $ \Lambda $.
Note that in implementing
the renormalization, a dimensionful parameter $\mu $ is introduced.
This is needed in order to specify the scale at which
the renormalization is done.  The upper limit on $\mu $
is fixed by the lattice spacing, so that the study of
lattice spacing corrections can be done with respect to a finite
$\mu $ as well as $ \Lambda $.

At this stage one must recall that our goal is to obtain the
universal scaling relations.  As such it does not matter what the
specific initial bare parameters are.  One is also
free to analyze the theory in terms of bare or renormalized
quantities.  To obtain the scaling relations the idea is,
having fixed the bare parameter at cutoff $ \Lambda $,
in order to maintain the renormalization conditions (2-7)
these parameters will develop a dependence on $ \Lambda $.  To determine
their behaviors one uses the fact that the renormalized theory
is independent of $ \Lambda $ when $ \Lambda \to \infty $.
Equivalently, one can demand that the bare theory be independent
of $\mu $.  The role of $\mu $ is similar to $ \Lambda $ in that
$\mu\to \infty $ corresponds to the limit of zero
lattice spacing.
Although the direct physical interpretation in terms of renormalized
quantities is obscured, it affords greater calculational ease
and accuracy.  Bearing in mind that even the bare theory is only
an effective coarse grained version of the original lattice theory,
not much is lost by this transcription.
However if one prefers one can always
keep in mind when making physical connections that the
renormalized parameters are related by equations (2-7), or in
different schemes by similar such, to the bare parameters
of the actual system.

Turning to the renormalization group equations, we will only
discuss them to the extent of making precise our physical discussion
above.  Keeping in the spirit of this conceptual presentation, our discussion
will be in
reference to the bare theory, where physical
contact is more transparent. For
a more detailed account, the reader is referred to the very careful
explanations given in \cite{amit,brezin1}.
The demand that the renormalized theory
be independent of the cutoff $ \Lambda $ gives,
$$
\left[\Lambda{\partial\over\partial\Lambda}+\beta(g_{0},\epsilon,
{\Lambda\over\mu}){\partial\over\partial{g_0}}-{N\over 2}\eta
(g_{0},\epsilon, {\Lambda\over\mu})\right]\Gamma^{(\mu)}(p_i,
g_{0},\Lambda)=0
\eqno(2-8)
$$
where
$$
\beta(g_{0},\epsilon,{\Lambda\over\mu})=\Lambda{d\over d\Lambda}
|_{g,\mu}g_0
\eqno(2-9a)
$$
$$
\eta(g_{0},\epsilon,{\Lambda\over\mu})=-\Lambda{d\over d\Lambda}
|_{g,\mu}{\ln Z}\left(g_0,{\Lambda\over\mu}\right)
\eqno(2-9b)
$$
The solution to which is,
$$
\Gamma^{(N)}(p,g_{0},\Lambda)=\phi\left(p,\ln \Lambda-
\int^{g_0}{dg^\prime\over\beta(g^\prime)}\right)\exp{N\over2}
\int^{g_0}{\eta(g^\prime)dg^\prime\over\beta(g^\prime)}
\eqno(2-10)
$$
The solution has the property
$$
\Gamma^{(N)}(p,g_0,\kappa\Lambda)=\Gamma^{(\mu)}(p,g_0(\kappa),\Lambda)
\exp{N\over2}\int_1^\kappa{d\gamma\over\gamma}\eta(g_0(\gamma))
\eqno(2-11)
$$
which implies the $ \Lambda \to \infty $ behavior is the same as studying
the flow of $g(\kappa)$ as $ \kappa\to \infty $.
To study the flow of $g$ requires the study of the beta-function(2-9a).
If $\beta $ has a zero, the asymptotic limit of $g$ as
$ \kappa\to \infty $ will be this zero provided $\beta ^\prime ( g^* )>0$.
For a general coupling $g \neq g^*$ the solution (2-10) has nontrivial
dependence on  the scale parameter.  On the other hand, at a fixed point,
using (2-11) one obtains,
$$
\Gamma^{(N)}({p\over \kappa},g_0,\Lambda)_{\kappa\to\infty}\to
\kappa^{-d+{1\over2}
(d-2+\eta)}\Gamma^{(N)}(p,g_0^*,\Lambda)
\eqno(2-12)
$$
thus attaining scale invariance.

Only at the fixed point is the theory scale invariant.  To
reach the fixed point requires either choosing the bare parameter
to be exactly $g^*_0$ or by taking the lattice spacing
to zero and moving along the renormalization flow
trajectory.  The former is highly unlikely whereas the
latter is physically only approximately possible since
statistical mechanical systems have a lattice cutoff.
As such, corrections to scaling will occur and some parts of it
are universal.  In particular universality only applies
to quantities computed at the fixed point.  On the other
hand, the imposition of a lattice, due to the finite
lattice spacing, implies this fixed point
is never reached under flow.  Deviations from exact
scaling are therefore reflected by deviations of the coupling
from
its fixed point value.  For small deviations an expansion
of the correlation functions about the fixed point
with respect to the coupling
constant can be
done. From this Taylor expansion, only the derivatives
of correlation functions evaluated at the fixed point are
universal.  This is the basic idea used to compute the
desired corrections.  In the next section we quantify
this discussion.

\bigskip
\noindent{\Large\bf Section 3}
\medskip

Our discussion of critical behavior will be in the context of
lattice gauge theory.  For this let us first
relate the O(4) theory to the effective QCD chiral Lagrangian
as hypothesized by \cite {pisarski}.
The order parameters of the effective chiral
Lagrangian were stated to be the quark condensates, which
can conveniently be written in matrix form as,
$$
M^i_j\equiv\langle\bar q^i_L q_{Rj}\rangle\eqno(3-1)
$$
The effective Lagrangian must be symmetric to
$SU(2)_L\times SU(2)_R\times
U(1)_{L+R}$ transformations.  This can be expressed by parametrizing
$M$ with ($\sigma $, $\pi $) and using the Pauli matrices to
get
$$
M=\sigma+i\bar\pi\cdot\bar\tau
\eqno(3-2)
$$
Using this form for $M$ in the effective Lagrangian, one can identify
four real independent fields with respect to which the theory is O(4)
symmetric.

Among the thermodynamic properties of the theory, the susceptibilities
for $T <  T_c $ are identified with the masses of the
$\sigma $-particle and pion.  There are two susceptibilities
since symmetry breaks only in one direction.  Defining $\sigma $ as the
symmetry breaking direction, $ m_\sigma^2 $ is the longitudinal
susceptibility,
$$
m^2_\sigma={\partial H\over\partial M}
\eqno(3-3a)
$$
and $ m_\pi^2 $ is the transverse susceptibility,
$$
m^2_\pi={H\over M}
\ . \eqno(3-3b)
$$

Let us now turn to the scaling behavior of the thermodynamic quantities
and their correction from scaling.  In the subsections to follow, we
will discuss the three types of scaling corrections that will occur.
These will be finite lattice corrections in subsection 3A, finite
volume effects in 3B and corrections from nonleading operator insertions
in 3C.  The former two give well prescribed effects that are not special
to the particular choice of lattice action.  An effective use of the latter
requires more specialized knowledge of the
lattice action.  The predictions
are still in the context of universal scaling effects.  However
the usefulness of an analysis of operator corrections
relies on first specifying an appropriate, manageable selection
of terms that dominate.

\medskip
\noindent{\bf 3A. Finite Lattice Spacing}
\medskip

To obtain corrections due to finite lattice spacing, as
discussed in Section 2, we must evaluate the solutions
of the renormalization group equations at $g$ away from $g^*$.
The basic idea is to Taylor expand the correlation
functions about the fixed point and retain only the
linear order deviation.  This introduces one new universal exponent,
$\omega = \beta^\prime (g^*) $,
first discovered by Wegner \cite {wegner}.
Details specific to our results below are given in appendix A.

With corrections included, one obtains the following
scaling behavior for the thermodynamic quantities \cite{brezin2,brezin3}.
Defining
$t= |T -  T_c | $, for $T> T_c $
the correlation length behaves as,
$$
\xi=A_\xi(a)t^{-\nu}(1+C_{\xi}L(a)t^{\omega\nu})
\eqno(3-4)
$$
the susceptibility as,
$$
\chi=A_\chi(a)t^{-\gamma}(1+C_\chi L(a)t^{\omega\nu})
\eqno(3-5)
$$
and the two point correlation function as,
$$
\Gamma^{(2)}(p,g,t,\mu)=A_2(a)p^{2-\eta}
\left[1+C_2L(a)t^{\omega\nu}+\tilde F(p)
\right]
\eqno(3-6)
$$

For the equation of state, $H(M)$ ,when $T< T_c $, there are two regions
to consider, $M/t^{1/\beta}\ll 1$ and $M/t^{1/\beta}\gg 1$.
In the first case we have
$$
H=A_H(a)M^\delta f\left({t_r\over M_r^{1\over\beta}}\right)
(1+C_H L(a)t^{\omega\nu})
\ , \eqno(3-7)
$$
where the Widom function $f(x)$ is defined in \cite{wallace},
so that the longitudinal and transverse susceptibilities
are respectively
$$
m^2_\sigma=A_m(a)m^2_{\sigma_0}(t_r,M_r)(1+C_H L(a)t^{\omega\nu})
\eqno(3-8a)
$$
$$
m^2_\pi=A_m(a) m^2_{\pi_0}(t_r,M_r)(1+C_HL(a)t^{\omega\nu})
\eqno(3-8b)
$$
where $ m_{\pi 0 } $ and $ m_{\sigma 0 } $ are the leading scaling
forms also given in \cite {wallace}.
In the latter case when $M/t^{1/\beta} \gg 1$, the equation of state is,
$$
H=A_{\tilde H}(a) M^\delta\left(1+C_{\tilde
H}L(a)M^{\omega\nu\over\beta}\right)
\eqno(3-9)
$$
The $r$ subscript on $t$ and $M$ in eqs. (3-7) and (3-8)
is to signify the ``rescaled''
temperature and
magnetization, which are defined in appendix A.  These shift the
absolute magnitude of $t$ and $M$ at different lattice
spacings but not the scaling behavior with respect to them
at a fixed lattice spacing.
In testing lattice corrections to scaling,
since the absolute magnitude of any one of the quantities
is generally fixed at each
lattice spacing by the leading scaling
term, the precise knowledge of the reduced quantities is
unneeded.
Note that the rescaling factors affect all the temperature and magnitization
variables above.  However, where possible we have absorbed these factors
into other coefficients, so as to leave the expressions in their most
convenient forms.
In the above the value of the critical
exponents are $\nu =0.73 \pm 0.02$,$\eta =0.03 \pm 0.01$,
$\beta =0.38 \pm 0.01$,
,$\delta =4.82 \pm 0.05$, $\gamma =1.44 \pm 0.04$, and $\omega \sim 0.46$.
All but the
latter are from a seven loop calculation in \cite {baker}
and the latter is computed from a large-n expansion
in \cite {brezin1,ma,bervillier}.

In the relations (3-4) to (3-9) the lattice dependence enters the correction
term as the factor $(g-g^*)\equiv L(a) t^{\omega\nu } $ where we have
explicitly
represented the dependence on the lattice constant, a, through
the nonuniversal function $L(a)$.  Recall that the coupling
constant $g$ is an artifact of the specific renormalization
scheme one adopts with no observable consequence.  However
since it enters in the same form, through the factor $(g-g^*)$
in the correction term, this implies the a-dependence of the
correction term  is reduced to the study of a single
function.  This is of relevance
since one can empirically determine $L(a)$ from a single measurement and
apply the so determined result to all other quantities.
For example it is possible to obtain a fit for
the function $L(a)$ from a given lattice measurement
at two different temperatures
of one of the thermodynamic quantities above.
Then by the predictions of the theory this determines the function $L(a)$
which controls the lattice dependence in the leading correction term
in all the other thermodynamic
functions.  An alternative approach is to vary the temperature $t$,
at fixed
lattice spacing and fit the measured  quantities to their respective forms
in eqs.(3-4) to (3-9).  One should then compute the ratio
between the coefficient of $ t^{\omega\nu } $ to that of the leading term,
for a given thermodynamic quantity, i, divided by the same
for any
other, j, which by the notation above is ${C_i}/{C_j}$. This
should be
the same at any lattice spacing $a$. In fact it has been shown
that these ratios are universal \cite {aharony}.
Either approach would determine
whether the predicted scaling corrections are there and
thus further test for the existence of a second order phase transition.

The definition of temperature T, in a lattice simulation requires some
careful consideration.  Recall that to vary temperature
in a lattice with a fixed number of time slices, one varies the
temporal lattice spacing.  This in turn is accomplished by varing
the coupling constant $\bar{\beta}=6/g^2$ which is related to the
lattice spacing through the $\beta$-function.  In such
a procedure, therefore, not only does the temporal but also spatial lattice
spacing change.  If one is sufficiently within the continuum
limit, the intrinsic short distance scale
would be independent of the lattice spacing and
$\bar{\beta}$, or the temperature associated with it, could
both serve as acceptable
temperature parameters.  This situation changes
greatly in the more realistic case when one has not reached
the continuum limit.  Now as $\bar{\beta}$ is varied, both the
temperature and correction coefficient, L(a),
change.  One remedy to this problem
is to use  different parameters in the temporal and
spatial directions.  If one is interested in
studying detailed properties of continuum
QCD, this is the most reasonable avenue to follow.
However, there is a way to considerably simplify the task
if all one is interested in is to establish the second order
nature of the transition.  If one accepts the assertion
that continuum QCD with two light quarks is in the same
universality class as the lattice-gauge action, then one
is free to study the critical behavior of the latter
in its own right without asking questions about
how its parameters relate to QCD.  In the context of
critical phenomena, this follows since
as far as the effective theory
of the two systems
is concerned, they differ at leading order in
operator corrections by only differences in their
respective short distance cutoffs.  Thus, although the
nonuniversal coefficients will differ, the leading
as well as nonleading universal scaling behavior with respect to temperature
is the same.  With this simplification, one is free to
use $\bar{\beta}$ as the temperature parameter of the lattice
theory.  Due to universality,
this provides a pragmatic method for verifying
scaling in the continuum theory also.


Observe that the scaling corrections vanish at the critical point,
$t=0$.  This is consistent with general notions about the second order
phase transition where at the critical point there is only
one length scale, that of the diverging correlation length,
and the system forgets about all its microscopic length scales.
Observe also that away from the critical point, $t \neq 0$, any
real many-body system will exhibit scaling corrections since there
is always a minimum microscopic length.


Within the theory of \cite {wilczek},
the two-point correlation function in (3-6) is supposed to be
for the $\pi $ and
$\sigma $ fields.  Recall that in the context of the lattice
theory, these fields were constructed from the fermion pseudoscalar
and scalar bilinears respectively.  In the theory of \cite {wilczek},
the relation (3-6) is supposed to reflect the coarse grained, long
distance
behavior of these fundamental
lattice correlators.  Analogously,
one can also write the two-point function for the $\rho $ and
$ a_1 $ fields which from \cite {rajagopal} are,
$$
\rho^i_{l}=\epsilon_{lmn}\phi_m\nabla^i\phi_n
\eqno(3-10a)
$$
$$
(A_1)^i_{l}=\phi_4\nabla^i\phi_{l}-\phi_l\nabla^i\phi_4 \ ,
\eqno(3-10b)
$$
The two-point correlation function for these fields is defined
as,
$$
\langle {O_i}^{\alpha\beta}(x){O_j}^{\gamma\delta }(0))\equiv
(\delta^{ij}-{{q^i q^j} \over q^2})
\Gamma^{(0,0,2)}\delta_{\alpha\beta},
\gamma\delta
\eqno(3-11)
$$
with ${O_i}^{\alpha\beta}\equiv
\epsilon^{\alpha\beta\gamma\delta}\phi^\gamma\nabla_i
\phi^\delta$, $1\leq\alpha,\beta,\gamma,\delta\leq 4$,
and has the scaling form with corrections,
$$
\Gamma^{(0,0,2)}(k,t,\mu)=c_0k^{d-2}g(kt^{-\nu})(1+c_{002}L(a)t^{\omega\nu})
\ . \eqno(3-12)
$$
The derivation of the correction factor also follows
from what is given in appendix A.
For these fields, accounting for corrections to scaling will
be more important than for the $\pi $ and $\sigma $ fields
since as pointed out by \cite {rajagopal}, mass terms dominate
the nonanalytic part except very close to the transition
point.  In practical terms this means scaling will be difficult to
detect so that maximal accuracy in the scaling form is desirable.
\eject
\bigskip
\noindent{\bf 3B. Finite Volume}
\medskip

Since all practical lattice studies are done on systems of finite
size,
studying the scaling behavior with respect to lattice
size can be a useful tool.
The renormalization group theory of finite size scaling
\cite {amit,brezin4,fisher} states that for any
multiplicatively renormalizable thermodynamical quantity, $P(t)$, such as
the correlation length or susceptibility,
$$
{P_L(t)\over P(t)}=g(L,t)
\eqno(3-13)
$$
where
$$
g(L,t)=F(L/\xi(t))
\ , \eqno(3-14)
$$
$P_L(t)$ is the value of the thermodynamic quality in the system of
finite size, and $\xi(t)$ is the correlation length
of the same system in the infinite volume limit.  Here $F(x)$ behaves as,
$$
F(x)\sim 1 \ \ \ \hbox{as} \ \ \ x\to\infty
\eqno(3-15a)
$$
$$
F(x)\sim x^{\rho/\nu} \ \ \ \hbox{as} \ \ \ x\to0
\ . \eqno(3-15b)
$$
where $\rho $ is the critical exponent associated with $P(t)$.

In a study which aims to test scaling behavior, the above can be
used to compare two systems at sizes $ L_1 , L_2 $,
without having to know the $L\to \infty $ behavior.  This can be implemented
as follows.  First measure, near the critical point,
a quantity, $ P_{L_i} ( t ) $,
as for example the susceptibility
$m(t)$, in systems one and two as a function of the temperature.
{}From this and using (3-15b), find the associated temperatures $ t_1 , t_2 $
such that
$$
\left({L_1\over L_2}\right)^{\rho/\nu}={P_{L_1}(t_1)\over P_{L_2}(t_2)}
\eqno(3-16)
$$
If the transition is second
order, then
for any other multiplicatively renormalizable thermodynamic
quantity, the same relation (3-16) should be found
at the same temperatures, except with $\rho $ replaced
by the critical exponent associated with the particular thermodynamic
quantity under examination.
Note that by this procedure, one never needs to
determine the precise critical
temperature in either of the two systems.  In eqs. (3-16), for notational
consistency, we used reduced temperatures on the right hand
side.  However, this relation can be
applied with knowledge only of the absolute temperatures.

\bigskip
\noindent{\bf 3C. Operator Corrections}
\medskip

The quartic O(4) theory (2-4) is only the limiting coarse
grained Hamiltonian obtained from the original lattice theory.
This limiting Hamiltonian retains the most relevant operators and
is exact for correlations in the zero momentum limit.
Away from this limit,
additive correction in an expansion in
$(p/ \Lambda $) arise for correlation functions at momentum $p$.
The explicit form of the corrections has both universal and
nonuniversal components.  The universal part arises from the
scaling behavior of the operator corrections.  The nonuniversal
part is the strength of each operator correction arising
in the effective theory.  The effective theory we are talking
about here is the one that interpolates between the exact lattice
action and the quartic LGW-Hamiltonian in (2-4).
We write the effective Hamiltonian as,
$$
H_{ e f f } = H_{ L G W } + H_c
\eqno(3-17)
$$
where $ H_{ L G W } $ is (2-4) and
contains the most relevant operators.  $ H_c $ has all the less
relevant operators which we write as,
$$
H_c = \sum^\infty_{j=1}\sum^{N_j}_{i=1}u_i^j O_i^j
\eqno(3-18)
$$
where $j$ is the canonical dimension of $ O_i^j $
and the sum over $i$ contains all operators in each set $j$.
For the two-point function , operator corrections lead to
the result,
$$
\Gamma_{cor}^{(2)}(p, g^*_0,\Lambda)=
\Gamma^{(2)}(p, g^*_0,\Lambda)+
\delta\Gamma^{(2)}(p, g^*_0,\Lambda)
\eqno(3-19)
$$
where
$$
\delta\Gamma^{(2)}(p, g^*_0,\Lambda)=p^2\left({p\over\Lambda}\right)^{-\eta}
\sum_\alpha u_\alpha\left({p\over\Lambda}\right)^{\eta_\alpha}
\eqno(3-20)
$$

The theory of operator corrections has been applied for a variety
of different
purposes in studies of
critical phenomena \cite {amit,brezin1,amit2,jug}.
For its relevance to lattice studies
at finite temperature, both lattice specific and
continuum QCD effects near the critical
point may be further understood by this procedure.
Such is an ambitious program which for one thing would need
outside model specific considerations
that limit the number of such correction terms to
a small set, amenable to numerical fitting.
By this decomposition one is able to systematically include
operator corrections based on their order of importance.

The dominant correction terms in $ H_c $ are from the operators
of canonical dimension six.  There are also dimension four operator
corrections but their effect has been accounted for in the finite
lattice spacing effects
of subsection 3A.  For the dimension six operators we have,
\setcounter{equation}{4}
\renewcommand{\theequation}{C\arabic{equation}}
\begin{eqnarray}
H_6 & = & {u^6_1\over
2!}(\nabla^2\bar\phi)^2+u^6_2\bar\phi\cdot\nabla^4\bar\phi+
	u^6_3\nabla^2(\bar\phi\nabla^2\bar\phi) \nonumber \\
& + & {u^6_4\over 2!}\nabla^4\bar\phi^2+{u^6_5\over3!}s_{ijkl}\phi_i\phi_j
	\phi_k \nabla^2\phi_l \\
& + & {u^6_6\over 4!}\nabla^2(\bar\phi^2)^2+{u^6_7\over 6!}(\bar\phi^2)^3
        \nonumber
\end{eqnarray}
The anomalous dimensions
they lead to are \cite {jug},
$$
\eta_1=\eta_2=\eta_3=0,
\eta_4={6\over8}\epsilon,\eta_5=\epsilon,
\eta_6=\epsilon,\eta_7=3\epsilon
\eqno({\rm C}6)
$$
The theory of operator corrections suggests a flexible method for the
analysis of critical behavior from lattice gauge simulations.
In particular it will be of use if more extensive
analysis of the fermion determinant is necessary.
Our discussion in this subsection was meant only to be
illustrative and to state its potential
relevance for lattice gauge studies.

\bigskip
\noindent{\Large\bf Conclusion}
\medskip

Model analogies and lattice studies have given plausible
evidence that the chiral transition is second order, but
as yet no proof.  To make progress, increased quantitative efforts
must be taken.  As we have seen in this paper, the renormalization-
group theory of critical phenomena provides an
ideal possibility.  For, not only do scaling studies
provide accurate information on second order behavior, but
they can be conducted in the realistic setting of lattice
specific complications.  What we have shown is that both finite
lattice spacing and volume effects imply well specified
correction terms to the scaling relations.  These corrections
involve both computable universal terms and nonuniversal terms that must
be fit to a small set of measurements.  Furthermore,
even if one could conclude from lattice gauge studies
that the $ N_f =2$ chiral transition
is second order, as indications show, it is
still necessary to know what the influence of lattice specific
effects are.  Since ultimately the goal of lattice studies
of finite temperature QCD is to provide accurate quantitative information
about various properties such as transition temperatures and the fermion
condensate, it is essential to understand the lattice specific effects.
It is well known by universality that the existence of a second
order phase transition occurs under very general
conditions, independent of most microscopic details.  Thus for lattice
studies only to establish that the transition is second order, although
a nontrivial first step, is still not enough.  The next
step is to assess the accuracy of the results and this
depends on the influence of the lattice, especially
the lattice spacing.

For this reason in many ways it is fortuitous that
the chiral $ N_f =2$ phase transition may  be second
order, since there is a well developed theoretical framework
from the renormalization group theory to assist in studying
these delicate issues.
In the present situation, where
there is no definitive
answer on the nature of the $ N_f =2$ transition
or even whether it is a transition or crossover behavior,
scaling theory provides a large set of consistency checks
to determine if it is second order.

The scenario hypothesized by \cite{pisarski,wilczek}
assumes that the fermionic
degrees solely determine the order parameter, neglecting any
explicit influence from gluonic degrees.  This is a
strong assertion that should be placed to stringent tests.
Although indications from both lattice studies and theoretical
models further support this outcome, they leave sufficient
uncertainties.  On the theoretical side, most calculations
already assume in some form that the order parameter is O(4)
symmetric and then proceed to compute a transition temperature
by some variant of a gap equation.  What is missing,
and no doubt nontrivial, is a derivation from QCD by way
of coarse graining, for example, in which the effective
O(4)-LGW action is derived.   Turning to lattice studies, here
technical uncertainties and results
that show strong dependence on the lattice spacing
\cite{gottlieb,engels,kapusta}
obscure
any definitive answers.

Since QCD is dynamically much richer than most statistical mechanical
systems studied on a lattice, further testing would be useful
in order to probe its refined properties.
With little guidance from analytic means,  the task is
placed on lattice studies to first determine whether there
is a transition or crossover effects and then obtained the detailed
features of whichever.  Although present day wisdom makes the
scenario of \cite {pisarski,wilczek} most plausible, as we stated above
the gluonic degrees play no explicit role.  However it
may be possible that Wilson-loop-like excitations,
such as glueballs perhaps, are sufficient in number
to play a role near the reported transition.  In such a case
it is possible that the appropriate
effective action which describes the loop dynamics is like the one
described by Polyakov \cite{polyakov}, which is also
a chiral invariant theory.  This could then
change the order parameters of the theory and the transition
to first order, fluctuation induced first order or
even eliminate the transition altogether.  Another possibility
discussed in \cite{pisarski} is that if the
instanton density just below $ T_c $ is
low enough so that $\eta^\prime $ becomes sufficiently
light, this would change
the order parameter from O(4) to O(2)$\times$O(4) and produce a
first order transition.

The above examples illustrate possible alternative interpretations,
which present data can not rule out.
To resolve between them and other such,
as a first step the predicted scaling from renormalization
group calculations should be tested against lattice simulation data.
Furthermore, to make an accurate analysis, it would be simultaneously
worthwhile to account for the effects of finite lattice spacing
and volume.  The former is especially important since it alters
the single exponent parametrization of scaling into two.

Apart from the fundamental issues regarding the nature of the
transition, accepting that it is second order, there are practical instances
where the lattice spacing has qualitative effects
that alter the continuum theory much as is already known
to be true at zero temperature
in usual lattice gauge studies.
In zero temperature studies the most common example is the violation
of chiral symmetry even for a massless theory.
In similar respect, aspects of the continuum behavior of systems
at finite temperature can be modified on a lattice.
To site an example of such a case,
the effect of finite lattice spacing
introduces dimension four operator corrections to the
effective action.  This in turn prevents a tricritical
point from occurring by a simple fine tuning as
conjectured by Wilczek \cite{wilczek}.
His claim is that the effect of a strange quark is to
renormalize operators of dimension four and higher.
Thus, he asserts that by
appropriately fine tuning its mass, it is possible to
make the quartic coupling vanish so leaving the the next
order, $ \phi^6 $ interaction, as the dominant one.  He then
notes that this theory gives a tricritical point.
Although lattice-model studies have been done on tricritical transitions
in the past with no problems, the scenario pictured by \cite {wilczek}
is slightly more subtle.  He envisions that the strange quark field
induces operator insertions of degree four and higher so as to
renormalize the original quartic coupling.
This effect from the strange quark field of renormalizing the couplings
is the same as will also occur from finite lattice spacing.
This means in the effective theory, both effects mesh together in redefining
the operator coefficients.

Although in the continuum theory where the lattice effects
are removed, there is no obvious problem with the so described scenario
of the strange quark,
if one
accounts for the lattice spacing,
quartic operator corrections from all sources
will hamper
the dimension six terms from controlling the critical behavior
in a manner as
simple as stated in \cite {wilczek}.  Nevertheless,
if the effects of the unwanted quartic terms can be made small,
by a small enough lattice spacing, it may be possible to
detect the crossover
towards the tricritical point.

Another place where finite lattice spacing effects are
relevant is in estimating the energy density in a volume
of order the correlation length near
$ T_c $.  With a finite lattice spacing,
the correlation length is modified by a factor
$(1+ C_{\xi} L(a) t^{\omega\nu } $).  Since $ t^{\omega\nu } $
tends slowly to zero near $ T_c $ these corrections
are potentially important and should be accounted for in
future estimates.  Also important are finite volume effects
not only for the usual technical
reasons, but also because they more closely simulate physical
reality in one
important practical applications, that of
heavy ion collision.

To summarize,
in this paper we have discussed the
use of the renormalization group theory of scaling corrections
in the context of lattice gauge theory.
A program based on the scaling relations given here would require
one empirically determined nonuniversal function
that depends
on the lattice spacing, which could be obtained from
fitting to a minimal set of measured thermodynamic
quantities.  Having done this, one would then have a complete set
of scaling relations that include
effects of finite lattice spacing.
These relations would provide one more cross check
for testing the consistency of simulation data.
For completeness it should be clear that we have discussed
only the leading correction from lattice spacing.  Higher order
corrections may also be important. This depends on what studies find
based on this simplist extension.



%

\bigskip
\noindent{\Large\bf Acknowledgements}

\medskip

\normalsize

I thank Professors Michael E. Fisher, Michael E. Peskin
and Douglas Toussaint
for helpful discussions and
comments in regards to this manuscript.
Financial support was provided
by
the U. S. Department
of Energy,
Division of High Energy and Nuclear Physics.

\bigskip
\noindent{\Large\bf Appendix A}
\medskip

In this appendix we will show that the lattice dependences in
scaling correction terms are of the forms given in (3-4)
to (3-9) and (3-12).  In particular we will establish that these correction
terms depend on the lattice through one nonuniversal
function $L(a)$.  Our derivation closely follows those given in
\cite{amit,brezin1}.  We hope our derivation below simplifies
theirs by being closer to the standard field theoretic language.
We will need correlation functions at temperature $t$ and magnetization
$M$.  For this we will have to be slightly more general than in
Section 2.  Following standard definitions, let
$ \Gamma^{(N,L)} $ denote the bare vertex function with L
insertions of $ \phi^2 $
and  N insertions of $\phi $.  This is related to the renormalized
correlation functions of the massless theory by,
$$
\Gamma_R^{(N,L)}(q_1,\ldots,q_N;p_1,\ldots,p_L;g,\mu,\Lambda)=
Z_{\phi^2}^L Z_\phi^{N/2}\Gamma^{(N,L)}
(q_1,\ldots,q_N;p_1,\ldots,p_L;g_0,\Lambda) \\ . \\
\eqno({\rm A}-1)
$$
Following \cite {amit} we take our renormalization conditions as,
$$
\Gamma_R^{(2,0)}(q,-q;g,\mu,\Lambda)|_{q^2=0}=0
\eqno({\rm A}-2a)
$$
$$
{\partial\over\partial p^2}\Gamma_R^{(2,0)}(q,-q;g,\mu,\Lambda)
|_{q^2=\mu^2}=1
\eqno({\rm A}-2b)
$$
$$
\Gamma_R^{(4,0)}(\{q\};g,\mu,\Lambda)|_{s.p.}=g\mu^\epsilon
$$
$$
\Gamma_R^{(2,1)}(q_1,q_2,p;g,\mu,\Lambda)|_{s.p.^\prime}=1
$$
$$
\Gamma_R^{(0,2)}(p,-p;g,\mu,\Lambda)|_{p^2=\mu^2}=0
$$
where $s.p.$ means in addition to
$q^2={3 \over 4} \mu^2$ that $q_i\cdot q_j={1\over4}\mu^2(4\delta_{ij}-1)$.
The point
$s.p.^\prime$ is where $q^2={3\over4}\mu^2,q_i \cdot q_j=-{1\over4}\mu^2$
so that
$p^2=(q_1+q_2)^2=\mu^2$.
We can construct correlation functions at finite $t$ and $M$ using the
ones above, which were for the symmetric, massless theory, as follows,
$$
\Gamma^{(N,L)}(q_i,p_i;t,M,g,\mu)=\sum^\infty_{I,J=0}{{M^I}{t^J}\over {I!J!}}
\Gamma_R^{N+I,L+J}(q_i,p_i;g,\mu)
\eqno({\rm A}-3)
$$
where $t$ is related to $ m_0 $ of the bare theory by,
$$
m^2_0=m^2_c+(m^2_0-m^2_c)=m^2_c+Z_{\phi^2} t
\eqno({\rm A}-4)
$$
so that $t\to0$ corresponds to the critical point.

Our task is to determine the scaling behavior of the correlation functions
in terms of the temperature and relate the correlation functions
to thermodynamic quantities.
The renormalized  correlation functions satisfy the renormalization
group equation,
$$
\left(\mu{\partial\over\partial\mu}+\beta(g){\partial\over\partial g}+{1\over2}
\eta (g)\left[N+M{\partial\over\partial M}\right]- \left[{1\over \nu(g)}-2
\right]\left[L+{\partial\over\partial t}\right]\right)
\Gamma_R^{(N,L)}(q,p,g,t,M)=0
\eqno({\rm A}-5)
$$
where $g=\mu^{-\epsilon}u$ is the dimensionless coupling constant and
$$
\beta(g)=-\epsilon\left({\partial\ln g_0\over {\partial g}}\right)^{-1}
$$
$$
{1\over \nu(g)}-2=\beta(g){\partial\ln Z_\phi\over \partial g}
$$
$$\eta(g)=\beta(g) {\partial\ln Z_\phi\over \partial g}
$$
which excludes the case $L=2$, $N=0$ at nonzero
magnetization, where an additional term is needed on the
right hand side.  This quantity will not be considered
in our work.   Above, $g_0$ is the dimensionless bare coupling constant,
which by dimensional analysis can only depend on g and the ratio
$\Lambda \over \mu$.  In the limit $\Lambda \rightarrow \infty$,
$g_0$ will therefore only depend on g.  This fact allows one to
express $\beta(g)$ in the form given above.  Further discussion
of this careful point is given in chapter 8 of \cite{amit}.

To obtain the homogeneous solution of (A-5),
one introduces a parameter
$ \lambda $ such that
$$
\ln\lambda=\int_g^{g(\lambda)}{dg^\prime\over\beta(g^\prime)}
\eqno({\rm A}-6a)
$$
with $g(\lambda=1)=g$ and ,
$$
\mu(\lambda)=\lambda\mu
\eqno({\rm A}-6b)
$$
$$
t(\lambda)=t\exp\left[-\int_g^{g(\lambda)}\{ {1\over\nu(g^\prime)}-2\}
{dg^\prime\over\beta(g^\prime)}\right]
\eqno({\rm A}-6c)
$$
$$M(\lambda)=M\exp\left[-{1\over2}\int_g^{g(\lambda)}{\eta(g^\prime)\over \beta
(g^\prime)}dg^\prime\right]
\eqno({\rm A}-6d)
$$
$$
\Gamma_R^{(N,L)}(q,p;g,t,M)
\eqno({\rm A}-6e)
$$
$$
=R^L(\lambda,g)S^N(\lambda,g)t_r^{-\nu
(d-N(d-2+\eta)/2)-L}\Gamma_R^{(N,L)}(qt_r^{-\nu},pt_r^{-\nu};g(\lambda),t=1,
M_{r}t_r^{-\nu(d-2+\eta)/2})
$$
from which one gets,
\setcounter{equation}{6}
\renewcommand{\theequation}{A-\arabic{equation}}
\begin{eqnarray}
\Gamma^{(N,L)}&&(q,p,g,t,M,\mu)=  \\
&&(\lambda\mu)^{d-2L-N(d-2)/2}
\left({M(\lambda)\over M}\right)^N\left({t(\lambda)\over t}\right)^L
\Gamma_R^{(N,L)}\left({q\over\lambda\mu},{p\over\lambda\mu};g(\lambda),
{t(\lambda)\over\lambda^2\mu^2}, {M(\lambda)\over(\lambda\mu)^{{d\over2}-1}},
\mu=1\right) \nonumber
\end{eqnarray}
where
$$
R(\lambda,g)=\exp\left[-\int^{g(\lambda)}_g\left({1\over\nu(g^\prime)}-
{1\over\nu}\right){dg^\prime\over\beta(g^\prime)}\right]
$$
$$S(\lambda,g)=\exp\left[-{1\over2}\int_g^{g(\lambda)}(\eta(g^\prime)-
\eta){dg^\prime\over\beta(g^\prime)}\right].
$$
Since $\lambda $ is arbitrary, we fix it such that $\lambda \to0$
corresponds to the critical region $t\to0$.
For this we make the standard choice
$$
{t(\lambda)\over \lambda^2\mu^2}=1
\eqno({\rm A}-8)
$$
so that from (A-6c) when $\lambda\to0 $ we have
$$
\lambda=\left[R(\lambda,g)t\right]^\nu\equiv t^\nu_r
\eqno({\rm A}-9)
$$
We want to examine the solution (A-7) for $g$ near $g^*$ in the direction
specified by the trajectory along $\lambda $.  Thus we Taylor expand
(A-7) with respect to $g$ as,
$$
\Gamma_R(g)=\Gamma_R(g^*)+(g(\lambda)-g^*){\partial\Gamma_R\over\partial g}
|_{g=g^*}
\ .\eqno({\rm A}-10)
$$
{}From (A-6a) and using $\beta(g(\lambda))=\omega(g(\lambda)-g^*)$
we find,
$$
{d\ln\lambda\over dg(\lambda)}={1\over \omega(g(\lambda)-g^*)}
\eqno({\rm A}-11)
$$
where $\omega\equiv\beta^\prime(g^*)$
is the Wegner exponent \cite{wegner}.
{}From this we have,
$$
g(\lambda)=g^*+(g-g^*)\lambda^\omega
\ . \eqno({\rm A}-12)
$$
Substituting into (A-10) we get the expansion,
\setcounter{equation}{12}
\renewcommand{\theequation}{A-\arabic{equation}}
\begin{eqnarray}
\Gamma_R^{(N,L)}(q,p;g,t,M,\mu) &&=  S^NR^Lt_r^{-\nu(d-N(d-2+\eta)/2)-L}
	\nonumber \\
&&\Gamma_R^{(N,L)}(qt^{-\nu}_r,pt^{-\nu}_r;g^*,t=1,SMt_r^{-\nu(d-2+\eta)/2},
\mu=1) \\
&& \times\left[
	1+(g-g^*)R^{\omega\nu}t^{\omega\nu}
\left\{
	\frac{
	\frac{\partial\Gamma_R}{\partial g}^{(N,L)}(qt_r^{-\nu},pt_r^{-\nu},g^*,1,
        SMt_r^{-\nu(d-2+\eta)}, \mu=1)
	}
	{\Gamma^{(N,L)}(qt_r^{-\nu},pt_r^{-\nu},g^*,1,
	SMt_r^{-\nu(d-2+\eta)},\mu=1)
	}
\right\}\right. \nonumber \\
&& \left.+ C_0\right] \nonumber
\end{eqnarray}
All lattice dependence is contained in $g$.  In particular this means
the correlation functions have an overall lattice dependent amplitude
factor $ R^N $ and a single function $L(a)\equiv g-g^*$
that enters as a factor in the leading scaling correction term.

What remains is to relate the correlation functions to the thermodynamic
quantities of interest.   The correlation length is defined as,
$$
\xi^2\equiv{\partial\Gamma_{bare}^{(2,0)}/\partial p^2|_{p^2=0}\over
\Gamma_{bare}^{(2,0)}|_{p^2=0}}={\partial\Gamma_{R}^{(2,0)}/\partial
p^2|_{p^2=0}\over\Gamma_{R}^{(2,0)}|_{p^2=0}}
\eqno({\rm A}-14)
$$
so that using (A-13) and suppressing unnecessary indices we get
\setcounter{equation}{14}
\renewcommand{\theequation}{A-\arabic{equation}}
\begin{eqnarray}
\xi^2&=&{
	{\partial\Gamma^{(2,0)}\over\partial p^2}(pt^{-\nu}_r,g^*)|_{p^2=0}\over
	\Gamma^{(2,0)} (pt^{-\nu}_r,g^*)|_{p^2=0}}+(g-g^*)R^{\omega\nu}t^{
\omega\nu}\nonumber \\
&\times&\left[{{\partial^2\Gamma^{(2,0)}(pt^{-\nu}_r,g^*)\over\partial
g\partial p^2}|_{p^2=0}\over\Gamma^{(2,0)} (pt^{-\nu}_r,g^*)}-
{{\partial^2\Gamma^{(2,0)}(pt^{-\nu}_r,g^*)\over\partial p^2}|_{p^2=0}\over
\left(\Gamma^{(2,0)}(pt^{-\nu}_r,g^*)\right)^2|_{p^2=0}}
{\partial\Gamma^{(2,0)}(pt_r^{-\nu},g^*)\over\partial g}|_{p^2=0}\right] \\
&=&R^{-\nu}t^{-2\nu}\left[1+(g-g^*)R^{\omega\nu}t^{\omega\nu}\left\{
{{\partial^2\Gamma^{(2,0)}(pt^{-\nu}_r,g^*)\over\partial
g\partial p^2}|_{p^2=0}\over{\partial\Gamma^{(2,0)}\over\partial p^2}
(pt^{-\nu},g^*)|_{p^2=0}}-
{{\partial\Gamma^{(2,0)}(pt_r^{-\nu},g^*)\over\partial g}|_{p^2=0}\over
\Gamma^{(2,0)}(pt^{-\nu}_r,g^*)|_{p^2=0}}
\right\}\right] \nonumber
\end{eqnarray}
which agrees with (3-4).
The susceptibility for $T> T_c $ is,
\setcounter{equation}{15}
\renewcommand{\theequation}{A-\arabic{equation}}
\begin{eqnarray}
\chi & \equiv &\Gamma_R^{(2,0)}(p=0;g,t) \\
& = & S^2t_r^{-\nu(d-2(d-2+\eta)/2)}
\Gamma_R^{(2,0)}(0;g^*,t=1,\mu=1)
\left[1+(g-g^*)R^{\omega\nu}t^{\omega\nu}
{
{\partial\Gamma_R^{(2,0)}\over\partial g}(0;g^*,1,\mu=1)
\over \Gamma^{(2,0)} (0;g^*,1,\mu=1)
}\right] \nonumber
\end{eqnarray}
which agrees with (3-5).
Finally the equation of state is,
\begin{eqnarray}
H(t,M,g) & \equiv & {{\partial\Gamma^{(0,0)}}\over{\partial M}}
=S{\partial\Gamma^{(0,0)}(t_rM_r,
	g^*)\over\partial g\partial M} + (g-g^*)R^{\omega\nu}t^{\omega\nu}
	\frac{\partial^2\Gamma^{(0,0)}(t_r,M_r,g^*)}
	{\partial g\partial M}\\
& = & SH(t_{r-1}, M_r,g^*)
	\left[1+(g-g^*)R^{\omega\nu}t^{\omega\nu}
{
{\partial^2 \Gamma^{(0,0)}(t_rM_r,g^*)\over\partial g\partial M}
\over
{\partial\Gamma^{(0,0)}(t_rM_r,g^*)\over\partial M}
}\right]\nonumber
\end{eqnarray}
where
$$
M_r=SM \ ,
$$
$$H(t_r,M_r,g^*)=f\left({t_r\over M_r^{1\over\beta}}\right)
$$
and the susceptibilities are,
$$
m^2_\sigma={\partial H\over\partial M}=S^2m^2_\sigma(t_r,M_r,g^*)
\left[1+(g-g^*)R^\omega t^{\omega\nu}
{
{\partial^3\Gamma^{(0,0)}\over\partial g\partial M^2}
(t_r,M_r,g^*)
\over
{\partial^2\Gamma^{(0,0)}\over\partial M^2}
(t_r,M_r,g^*)
}\right]
\eqno({\rm A}-18b)
$$
$$
m^2_\pi={ H\over M}=S^2m^2_\pi (t_r,M_r,g^*)
\left[1+(g-g^*)R^\omega t^{\omega\nu}
{
{\partial^2\Gamma^{(0,0)}\over\partial g\partial M}
(t_r,M_r,g^*)
\over
{\partial\Gamma^{(0,0)}\over\partial M}
(t_r,M_r,g^*)
}\right]
\eqno({\rm A}-18a)
$$

Since the parameter $ \lambda $ in our analysis was arbitrary,
above we conveniently chose it to study scaling behavior
with respect to the temperature $t$.
In a similar way, one can study scaling with
respect to the magnetization $M$ in the equation of state.  In particular
when $x\to0$, so that $M\gg t$,  we have $f(x)\sim1$ and the $M$-dependent
scaling correction term dominates over the $t$-dependent one.
To examine this we chose $ \lambda $ such that
$ (M(\lambda))/((\lambda\mu)^{d/2-1}=1$
which implies
$$
\lambda=\left({SM\over\mu^{{d\over2}-1}}\right)^{2\over d-2+\eta}
\eqno({\rm A}-19)
$$
Substituting into (A-13) we find,
$$
H=(SM)^\delta\left(1+(g-g^*)\left({SM\over \mu^{{d\over2}-1}}\right)^{\omega\nu
	\over\beta}
{{\partial^2\Gamma^{0,0)}\over\partial g\partial M}
(t_r,M_r,g^*)
\over
{\partial\Gamma^{0,0)}\over\partial M}
(t_r,M_r,g^*)}
\right)
\eqno({\rm A}-20)
$$

\medskip
*  Present address: Department of Physics, the Pennslyvania State University,
104 Davey Laboratory, University Park, PA 16802-6300


\medskip
\begin{thebibliography} {99}
\bibitem{pisarski} R. Pisarski and F. Wilczek,  Phys. Rev. D29, 338
(1984).
\bibitem{wilczek}
F. Wilczek, Int. J. Mod. Phys. A7, 3911 (1992).
\bibitem{rajagopal} K. Rajagopal and F. Wilczek,  Nucl. Phys. B399, 395
(1993).
\bibitem{karsch} F. Karsch, HLRZ-93-62, Sept. 1993.
\bibitem{amit} D. J. Amit, {\it Field Theory, the Renormalization
Group, and Critical Phenomenon}, World Scientific, (1984).
\bibitem{brezin1} E. Brezin, J. C. Le Guillou, J. Zinn-Justin in
{\it Phase Transitions and Critical Phenomena 6, 125 (1976)},
ed. C. Domb and J. Green (Academic Press).
\bibitem{wegner} F. J. Wegner, Phys. Rev. B5, 4529 (1972);
Phys. Rev B6, 1891 (1972).
\bibitem{brezin2} E. Brezin, J. C. Le Guillou, J. Zinn-Justin,
Phys. Rev D8, 434 (1973).
\bibitem{brezin3} E. Brezin, J. C. Le Guillou, J. Zinn-Justin,
Phys. Rev D8, 2418 (1973).
\bibitem{wallace} D. J. Wallace in
{\it Phase Transitions and Critical Phenomena 6, 293 (1976)},
ed. C. Domb and J. Green (Academic Press).
\bibitem{ma} S. K. Ma, Phys. Rev. A10, 1818 (1974).
\bibitem{bervillier} C. Bervillier, G. Girardi, and E. Brezin,
Saclay preprint (1974).
\bibitem{baker} G. Baker, B. Nickel, and D. Meiron, Phys. Rev. B17,
1365 (1978).
\bibitem{aharony} A. Aharony and M. E. Fisher, Phys. Rev. Lett. 45,
679 (1980).
\bibitem{brezin4} E. Brezin, Journal de Physique (Paris) 43, 15 (1982).
\bibitem{fisher} M. E. Fisher and V. Privman, Phys. Rev. B32, 447
(1985).
\bibitem{amit2} D. J. Amit, D. J. Wallace and R. K. P. Zia,
Phys. Rev B15, 4657 (1977).
\bibitem{jug} G. Jug, Ann. Phys. 142, 140 (1982).
\bibitem{gottlieb} S. Gottlieb, W. Liu, R. L. Renken,
R. L. Sugar, and D. Toussaint, Phys. Rev. D38, 2888 (1988).
\bibitem{engels} J. Engels, S. Karsch, and H. Satz, Nucl. Phys.
B205, 239 (1982).
\bibitem{kapusta} J. Kapusta, Physica A158, 125 (1989).
\bibitem{polyakov} A. M. Polyakov, {\it Gauge Fields and Strings},
(Harwood Academic Publishers), chapter 7.
\end{thebibliography}
\end{document}